\documentclass[aps,prl,reprint,groupedaddress,showkeys,nofootinbib]{revtex4-1}

\usepackage{amssymb}
\usepackage{ifpdf}
\ifpdf
	\usepackage[pdftex]{graphicx}
	\usepackage{epstopdf}
\else
	\usepackage[dvipdfm]{graphicx}
\fi
\usepackage{natbib,hyperref}

\begin{document}

\title{Photon subtraction from traveling fields\\
- recent experimental demonstrations}

\author{Jonas S. Neergaard-Nielsen} \author{Makoto
Takeuchi} \author{Kentaro Wakui} \author{Hiroki Takahashi$^{\dagger}$}
\author{Kazuhiro Hayasaka} \author{Masahiro Takeoka}
\author{Masahide Sasaki} \affiliation{National Institute of
Information and Communications Technology (NICT)\\ 4-2-1
Nukui-kitamachi, Koganei, Tokyo 184-8795, Japan\\
$^{\dagger}$Present address: University of Sussex, Pevensey 2, Falmer,
Brighton,
  East Sussex, United Kingdom BN1 9QH}

\begin{abstract}
%We review our most recent results on application of the photon
%subtraction technique for different optical quantum information
%processing primitives, along with a brief summary of other experimental
%accomplishments in the field.
We review our most recent results on application of the photon
subtraction technique for optical quantum information processing
primitives, in particular entanglement distillation and generation of
squeezed qubit states. As an introduction we provide a brief summary of
other experimental accomplishments in the field.
\end{abstract}


\maketitle

\section{Introduction}

For a number of years, our group has been working on generation,
manipulation and detection of non-classical quantum states of light
using techniques and concepts from both the continuous variable (cv)
and the discrete variable (dv) subfields of quantum information
science. By doing so, we have taken part in a trend -- both theoretical
and experimental -- of moving away from seeing the cv and dv paradigms
as separate, competing frameworks for quantum information processing
and rather look for ways of taking advantage of and combining the best
properties of both.

In this paper, we review our two most recent results in this direction:
Entanglement distillation of Gaussian states \citep{Takahashi2010a} and
generation of arbitrary qubit-like squeezed states
\citep{Neergaard-Nielsen2010a}. These two experiments demonstrate some
of the possibilities that arise when combining squeezed vacuum or other
non-classical cv states with the technique of photon subtraction.
Before getting to that, we give a short (and incomplete) overview of a
number of results in quantum optics and quantum information that
brought us to the current state of affairs.

\section{A brief history of bridges}

\subsection{Motivation}

When optical quantum information processing was developed through the
90's, a reasonably strong distinction was made between the two
different ``regimes'' of discrete and continuous variable systems. This
distinction persists to some extent even today. In fact there is no
fundamental difference between the two; rather, they are complementary
ways of describing the states of quantized electromagnetic fields
\citep{andersen2010}.

Most quantum optics experiments and quantum information schemes
consider information contained in discrete degrees of freedom of an
optical field: photon numbers, orthogonal polarization or spatial
modes, time bins and so on. The research of a number of groups,
however, looks into states of light that are more conveniently
expressed in terms of continuous degrees of freedom, typically as
non-commuting observables like the phase and amplitude quadratures or
Stokes polarization vectors. A typical state representation in dv is
the density matrix over the available orthogonal modes of the system,
while cv states are often represented by their Wigner functions --
quasi-probability distributions in the phase space of conjugate
variables. However, a corresponding density matrix can be
straightforwardly calculated from a Wigner function and vice-versa, so
the choice of representation is just one of convenience, clarity and
perhaps convention.

The interest in using continuous variables for quantum information
processing originates largely from the efficiency of various processes:
Squeezed light and coherent states can be generated deterministically,
homodyne detection can achieve almost 100\% efficiency, and many
standard operations are deterministic and easy to implement in the lab.
Furthermore, all of these processes can be described in the simple and
well understood Gaussian framework where states are represented solely
by their means and covariance matrices. Cv quantum information
processing is not limited to optical systems, but have also been
explored in atomic systems such as spin-polarized ensembles of neutral
atoms, and there are several promising approaches for efficiently
implementing light-atom interactions \citep{Hammerer2008}. Finally, it
should be mentioned that modern optical communications uses quadrature
encoding (cv) and therefore cv approaches to quantum information may
yield better compatibility with existing infrastructure.

Experimentally, one of the driving forces behind optical
implementations of quantum information primitives and protocols has
been the process of parametric down-conversion. Although there exist
protocols for quantum key distribution using only classical coherent
states, for basically all other purposes a source of non-classical
light is required, and for that, parametric down-conversion is one of
the simplest and most efficient methods around. By a second-order
nonlinear interaction in a crystal or waveguide, a single photon from a
pump beam is converted into two daughter photons correlated in a number
of different degrees of freedom. For weak pumping, this results in
generation of photon pairs, often with orthogonal polarization (type II
down-conversion) such that they are easily separable. For stronger
pumping and degenerate down-converted photons, the single output beam
has quantum noise fluctuation below the standard quantum limit and is
said to be in a squeezed state. These two different aspects of the
down-conversion output lend themselves to detection either by photon
counting/single photon detection or by homodyne detection,
respectively. Photon detection gives a naturally discrete measurement
outcome, while homodyne detection provides a continuous-valued
measurement of a given quadrature variable. While the photon generation
and measurement process is probabilistic, squeezing generation and
homodyne measurement is fully deterministic. This was reflected in the
first experimental demonstrations of quantum teleportation by
\citet{Bouwmeester1997} (dv) and \citet{Furusawa1998} (cv) -- these
almost simultaneous papers are perhaps the most prominent examples of
the distinction between the two paradigms.

\subsection{Experiments}

In 2001, \citet{Lvovsky2001} presented probably the first experiment to
truly bridge the dv and cv worlds. They generated a single photon state
by heralding on detection of one of the photons of a parametrically
down-converted twin pair, and then measured its Wigner function by
homodyne tomography (full state characterization by homodyne detection
of a large ensemble of identically prepared states). They essentially
measured the continuous-variable representation of an intrinsically
discrete quantum state, the one-photon Fock state. Its Wigner function
showed a prominent dip below zero at the phase space origin -- a
feature that is common to all non-Gaussian pure states, but that had
not been observed in optics until then, since homodyne measurements had
only been applied to Gaussian states such as squeezed vacuum or
coherent states. This accomplishment was followed up by
\citet{Babichev2004} who performed two-mode tomography of a dual-rail
single photon, and by \citet{Zavatta2004b} who showed the Wigner
functions of photon-added coherent states, obtained by seeding a
coherent beam into one output of a non-degenerate parametric
down-converter.

A later breakthrough in the merger of dv and cv technologies came when
\citet{Ourjoumtsev2006c, Neergaard-Nielsen2006a} and later
\citet{Wakui2007a} combined squeezed vacuum generation with photon
detection-heralded homodyne tomography to demonstrate small coherent
state superposition states, also called Schr{\"o}dinger kittens. Such
states of the form $|\psi\rangle = |\alpha\rangle -
\left|-\alpha\right\rangle$ will be useful for quantum information
processing \citep{vanEnk2001, Gilchrist2004a, Jeong2007a, Lund2008a}
but are unfortunately extremely difficult to prepare in a deterministic
manner. \citet{Dakna1997a} suggested already in 1997 a way to
probabilistically make high-fidelity approximations to these states for
small coherent amplitudes $\alpha$, simply by subtracting one or more
photons from squeezed vacuum. \citet{Wenger2004} showed the first signs
of non-Gaussian statistics from a state prepared in this way, and this
was later followed up by the full negative Wigner function state
preparations \citep{Ourjoumtsev2006c, Neergaard-Nielsen2006a,
Wakui2007a}. The photon subtraction operation itself has turned out to
be a generally useful tool for quantum information processing and for
fundamental tests of quantum mechanics, as explored in a series of
theoretical studies \citep{Opatrny2000, Browne2003, Nha2004,
Garcia-Patron2004, Jeong2008a}.

These experiments originate in the cv world with squeezing and homodyne
tomography, but incorporate photon detection and conditional state
preparation from the dv toolbox. In the measured output states, their
dual nature is reflected in the Wigner function which is a mix between
a single photon state and a squeezed vacuum (it is in fact a squeezed
photon). We can therefore no longer claim convincingly that the
generated states or the experiments as such belong to the dv or the cv
regimes -- rather they are bridging these two worlds. This trend of
dv/cv bridging experiments that slowly started with the results from
Lvovsky and Bellini's labs in the first half of the decade has taken
off dramatically in the last few years, especially since the first
kitten generation demonstrations in 2006-7. To our knowledge, kitten
states have now been observed in Paris, Copenhagen, three labs in
Tokyo, and at NIST. Most recently, photon-number resolving transition
edge sensors were used to prepare three-photon subtracted squeezed
vacuum \citep{Gerrits2010a} and telecom wavelength two-photon
subtracted squeezed vacuum \citep{Namekata2010}.

Other interesting experiments that are worth mentioning are the
two-photon Fock state tomography \citep{Tualle-Brouri2006}, the
time-delayed two-photon subtraction for enlarging the even kitten state
$\left|\alpha\right\rangle + \left|-\alpha\right\rangle$
\citep{Takahashi2008a}, a squeezed kitten state prepared by conditional
homodyning of a 2-photon Fock state \citep{Ourjoumtsev2007c}, remote
preparation of entangled kitten states
\citep{Ourjoumtsev2009Preparation}, a direct test of the commutation
relations, confirmed by homodyne tomography \citep{Zavatta2009}, a
tunable POVM detector combining homodyne detection with photon number
resolution \citep{Puentes2009}, tomography of photons from a pulse
pumped OPO \citep{Nielsen2009}, tomography of 0-, 1-, 2-photon Fock
state superpositions \citep{Bimbard2010} and of intermediate single
photon/squeezed vacuum states \citep{Jain2010}, noiseless amplification
of weak coherent states by photon detection \citep{Zavatta2010,
Xiang2010, Ferreyrol2010, Usuga2010}, photon counting-assisted optimal
coherent state discrimination \citep{Wittmann2008, Tsujino2010,
Wittmann2010}, and teleportation of a kitten state \citep{Lee2009}.
\citet{Kim2008c} has also compiled a nice review of the theoretical
foundations and experimental progress in this research area, while
\citet{Lvovsky2009} gives a comprehensive review focused on the methods
and applications of cv state tomography. A more general overview of
various dv/cv hybrid approaches to quantum information processing is
provided by \citet{VanLoock2010}.

\section{Experiment basics}

Both of the two experiments detailed later are based on our setup for
generation of photon-subtracted squeezed vacuum (kitten state). We will
here briefly go through some technical details of this experiment, so
that we can better focus on the conceptual ideas later.

\begin{figure}
  \includegraphics[width=\columnwidth]{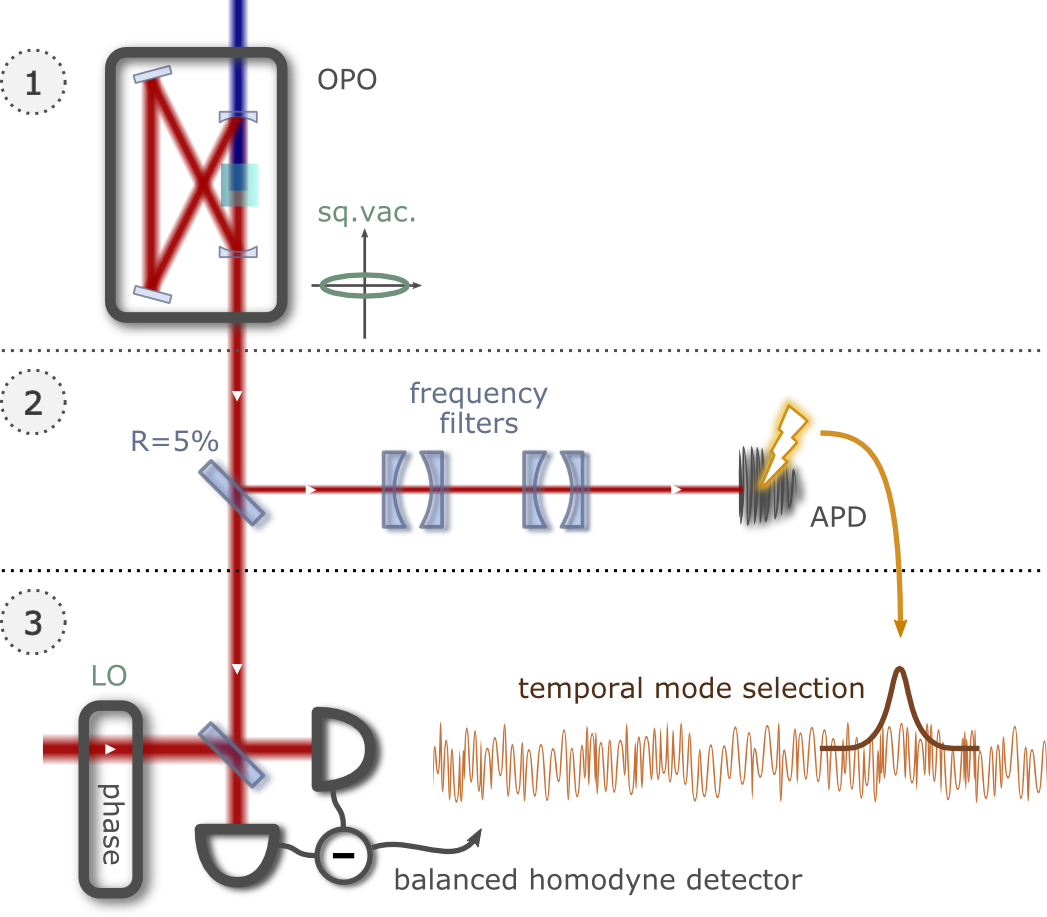}\\
  \caption{Conceptual setup for generation of single-photon subtracted
  squeezed vacuum. 1) Squeezed vacuum preparation. 2) Photon
  subtraction. 3) Homodyne tomography.\label{fig:setupcat}}
\end{figure}

There are essentially three stages of the kitten generation: Resource
state preparation (squeezed vacuum), state manipulation (photon
subtraction) and state characterization (homodyne tomography). These
are illustrated in Figure \ref{fig:setupcat} and described below in
turn. For more details, refer to \citep{Wakui2007a, Takahashi2008a}.

\subsection{Squeezed vacuum}

Our source of non-classical light is a continuously pumped optical
parametric oscillator (OPO) consisting of a 10 mm long periodically
poled potassium titanyl phosphate (PPKTP) crystal in a bow-tie type
single-ended optical cavity. It is pumped by a frequency-doubled
Ti:Sapph laser with pump wavelength 430 nm. The cavity has a 580 MHz
free spectral range and the FWHM bandwidth is 9 MHz, and because of the
low internal losses, the outcoupling efficiency is as high as 96\%,
which leads to almost pure squeezing at the output. This OPO
configuration has proven to be able to deliver up to 9 dB of squeezing
\citep{Takeno2007}, but high squeezing levels result in a higher degree
of mixedness in the observed temporal modes because the spectral
character of the state becomes ``skewed''; the bandwidth of the
squeezed quadrature is $\gamma+\epsilon$, while for the anti-squeezed
quadrature it is $\gamma-\epsilon$, where $\gamma$ is the OPO bandwidth
and $\epsilon$ the pump parameter \citep{Molmer2006,Takeoka2010}. For
this reason, we usually generate squeezing in the range 2--5 dB.

In order to control the orientation of the squeezing in phase space, we
inject a coherent probe beam through one of the high-reflectivity
mirrors of the OPO. This beam then co-propagates with the squeezed
vacuum and can be used for phase locking as well as locking of the
filter cavities in the photon subtraction channels. For phase locking
the squeezing orientation, we monitor the transmitted beam on a
standard DC detector. The beam has undergone classical phase-dependent
parametric amplification in the OPO, so by fixing the amount of
amplification, the probe beam will be locked in phase with the squeezed
vacuum. By later locking the probe beam to the local oscillator of the
homodyne detector (which determines the phase space orientation), we
can fix the squeezing orientation.

The probe beam is much stronger than the squeezed vacuum, and since the
photon detector (APD) is frequency undiscriminating and cannot
distinguish between the narrow-band probe and the wide-band squeezing,
we need to avoid the probe beam when measuring the kittens. Therefore
we introduced a chopping cycle of the probe beam where it is on for 20
$\mu$s and off for the next 80 $\mu$s, chopped by two double-pass AOMs.
While the probe is on we do the phase and cavity locking and shield the
APD. While it is off, we uncover the APD and measure the conditional
homodyne signals. In the meantime, the locks are kept passively stable.

\subsection{Photon subtraction}

The squeezed light impinges on a 5\% reflection beamsplitter, where the
main transmitted part goes directly to the homodyne detector for output
state analysis. The small reflected part is directed to an avalanche
photo diode (APD), whose click signal is connected to a digital
oscilloscope to act as a conditioning trigger of the homodyne data
acquisition. The squeezed light does not contain many photons, only 5\%
is tapped for trigger detection, and the total transmission and
detection efficiency of the trigger channel is only about 10\%, so the
photon count rate is rather low -- around 10,000 counts/sec for 3 dB
squeezing. But whenever a photon is detected, we know that it is
missing from the correlated part of the main signal beam, which is
therefore projected to a photon-subtracted squeezed vacuum state. The
low efficiency of the trigger channel only influences the success rate
of the experiment; the impact on the generated states is almost
negligible.

An APD and a homodyne detector have completely different spectral and
temporal mode resolutions, so the photon detected by the APD is not
necessarily correlated with the mode seen by the homodyne detector.
That is true for both a pulsed laser setup \citep{Ourjoumtsev2006c} and
for a cw setup \citep{Neergaard-Nielsen2006a, Wakui2007a}, but the
problem is manifested in different ways and requires different
solutions. In the cw case, the down-conversion process happens over a
nanometer-broad spectral range, but only at those frequencies resonant
with the OPO cavity. The APD can detect all the down-converted photons,
but the homodyne detector is limited to a bandwidth of around 50 MHz
around the carrier frequency of the local oscillator, derived from the
main 860 nm laser. Therefore we have to eliminate all the
non-degenerate OPO spectral cavity modes from the light reaching the
APD. We do that by passing the light through two subsequent Fabry-Perot
cavities, both resonant with the degenerate OPO mode but with
different, rather wide free spectral ranges. This arrangement works
very well and seemingly filters away all unwanted photons.

\subsection{Homodyne tomography}

To measure the Wigner functions of the generated non-classical light
states, we interfere them with a strong local oscillator (LO) on a
50:50 beam splitter, detect the two output beams on fast PIN photo
diodes and monitor the amplified difference photocurrent on the
oscilloscope triggered by the APD signal. For every trigger click, we
record a trace of about 1 $\mu$s length around the trigger time at a
sample rate faster than twice the detector bandwidth. Later, in
post-processing, we perform a temporal mode selection by weighting each
trace with a mode function
\begin{equation}
f(t) = \kappa e^{-\gamma|t-t_0|} - \gamma e^{-\kappa|t-t_0|} ,
\end{equation}
where $\gamma$ is the OPO HWHM bandwidth, $\kappa$ the filter cavity
bandwidth, and $t_0$ the time of photon detection. This mode function
corresponds to the temporal extent of the correlations between the
detected photon and the photon-subtracted main beam, so it is close to
the optimal temporal filter for extracting the conditionally prepared
state with high efficiency.

\begin{figure}
  \includegraphics[width=\columnwidth]{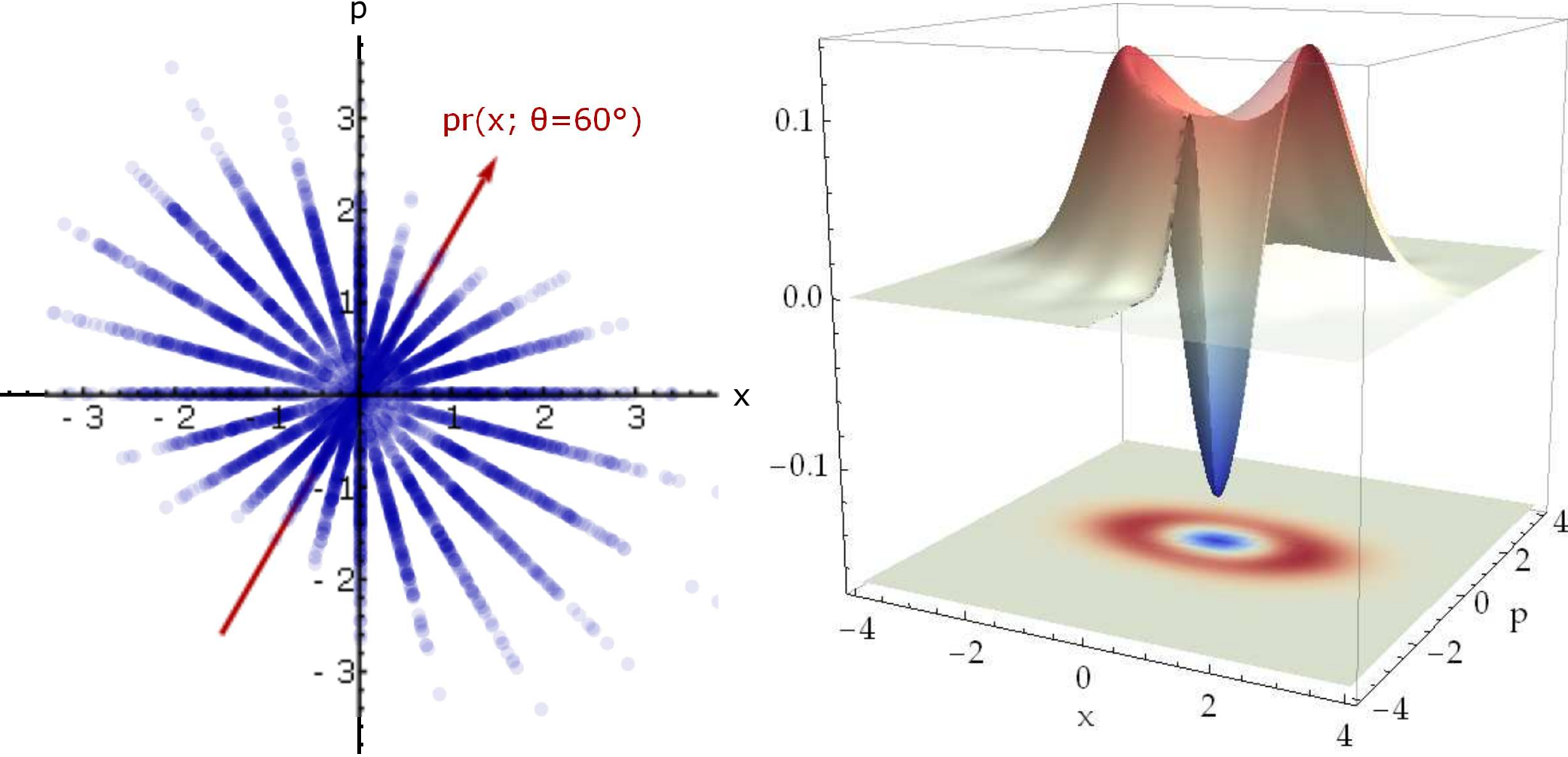}\\
  \caption{Example of homodyne tomography of a single-photon subtracted
  squeezed vacuum state. On the left is a series of homodyne
  measurements taken at 12 different fixed phase space angles. Notice
  the relative lack of observations around the origin -- this is
  reflected in the negative dip of the reconstructed Wigner function in
  the right panel. The Wigner function has been rotated in phase space
  to have squeezing along the p-axis.\label{fig:tomography}}
\end{figure}

The homodyne detection process described here results in a single
outcome of the quadrature measurement at the given LO phase. To
reconstruct the full state, we repeat the measurement many thousand
times for identically prepared states, with the LO phase locked to one
of 6 or 12 evenly distributed values between 0 and $\pi$. Finally, we
process the total set of quadrature measurements with a maximum
likelihood algorithm \citep{Lvovsky2004, Lvovsky2009} to estimate the
density matrix of the generated state. An example of a full tomographic
measurement along with the reconstructed state (Wigner function
calculated from the density matrix) is presented in Figure
\ref{fig:tomography}.

\section{Distillation of Gaussian entanglement}

\begin{figure}
  \includegraphics[width=\columnwidth]{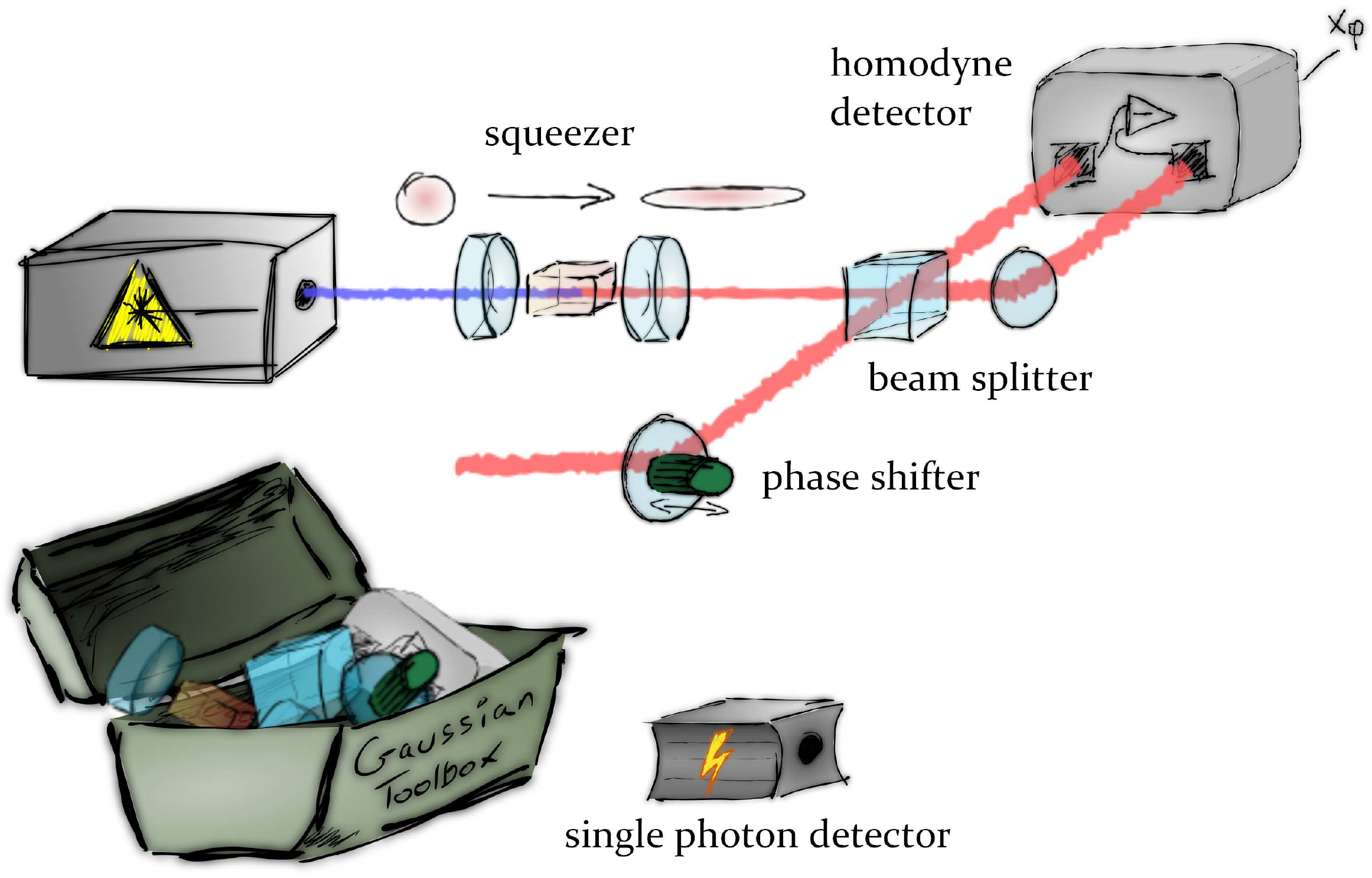}\\
  \caption{The Gaussian toolbox. A basic continuous-variable experiment
  involving the essential Gaussian operations. For a number of
  important quantum information tasks, this set of Gaussian tools is
  not sufficient and must be supplemented by a non-Gaussian tool such
  as a single photon detector.\label{fig:toolbox}}
\end{figure}

One of the great advantages of cv quantum information processing is
that it is essentially constructed from deterministic, Gaussian
operations, that is, operations with Hamiltonians that are at most
quadratic in the field operators. All of the most commonly used
techniques in the lab such as squeezing, phase shifting, beam
splitting, displacement and homodyne detection (Figure
\ref{fig:toolbox}) are Gaussian operations, which means that an input
state with Gaussian Wigner function (or characteristic function)
remains Gaussian after the transformation. The theory for Gaussian
states, entanglement, operations and so on is very well developed and
provides a powerful framework for various quantum information
processing tasks \citep{Loock2005a, Cerf2007}. Unfortunately, there are
some significant exceptions to what is possible with Gaussian
resources. \citet{Bartlett2002} showed that a quantum computer using
only Gaussian resources can be efficiently simulated by a classical
computer. Therefore non-Gaussian operations are needed to achieve a
quantum-mechanical speed-up of computing. Another problem is
long-distance distribution of entanglement which is needed for quantum
communication protocols. In order to combat propagation losses, the
entanglement must be reinforced by quantum repeater stations
\citep{Briegel1998a}, incorporating entanglement swapping and
distillation \citep{Bennett1996}. However, it was shown that a Gaussian
entangled state cannot be distilled by Gaussian local operations and
classical communication \citep{Fiurasek2002,Eisert2002,Giedke2002}.
Since Gaussian entanglement is the predominant type of cv entanglement
(such as two-mode squeezed EPR states), non-Gaussian operations are
required for distribution over long distances.

\citet{Hage2008, Hage2010} and \citet{Dong2008} performed entanglement
distillation using only conditional homodyne detection, but in those
cases, the Gaussian entangled beams had been exposed to non-Gaussian
noise. Then it is possible to recover the original amount of Gaussian
entanglement by Gaussian operations such as homodyne detection.
However, the probably most common and critical transmission noise comes
from linear loss, which is a Gaussian transformation.
\citet{Ourjoumtsev2007b} demonstrated that a non-local (collective)
photon subtraction can increase entanglement that is still Gaussian.
Based on this demonstration and on the realization by
\citet{Opatrny2000} that photon subtraction can increase entanglement
and lead to better teleportation, we performed the first entanglement
distillation \footnote{We note that there is some confusion about
terminology in the literature about entanglement distillation: By some
accounts, our result should rather be classified as entanglement
\emph{concentration}, while the term \emph{distillation} is in line
with other sources -- notably the conceptually similar procrustean
method for qubit distillation of Kwiat \emph{et al.}, Nature
\textbf{409}, 1014 (2001).} of a Gaussian two-mode state
\citep{Takahashi2010a}.

\subsection{Experiment}

\begin{figure}
  \includegraphics[width=\columnwidth]{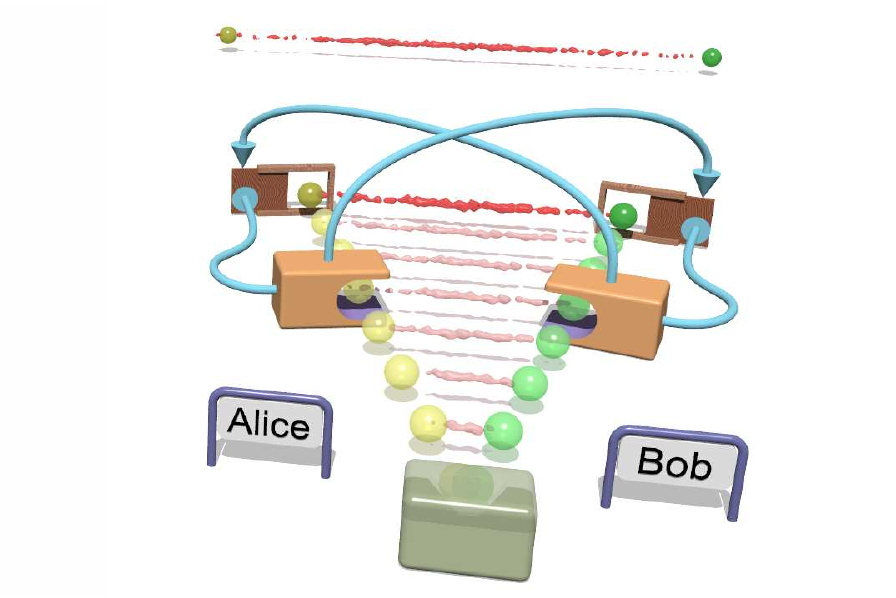}\\
  \caption{Basic concept of entanglement distillation. Weak
  entanglement is enhanced by probabilistic local operations (orange
  boxes) followed up by classical communication (blue wires) to select
  those entangled pairs that have been successfully distilled.
  \label{fig:distillation-concept}}
\end{figure}

The basic concept of the distillation protocol, illustrated in Figure
\ref{fig:distillation-concept}, is to prepare a series of entangled
states shared by Alice and Bob, let them perform local operations on
their part and finally let them communicate with each other classically
to determine which states have successfully had their entanglement
increased and therefore can be kept for further processing. This
process sacrifices the rate of shared states for the amount of
entanglement.

\begin{figure}
  \includegraphics[width=\columnwidth]{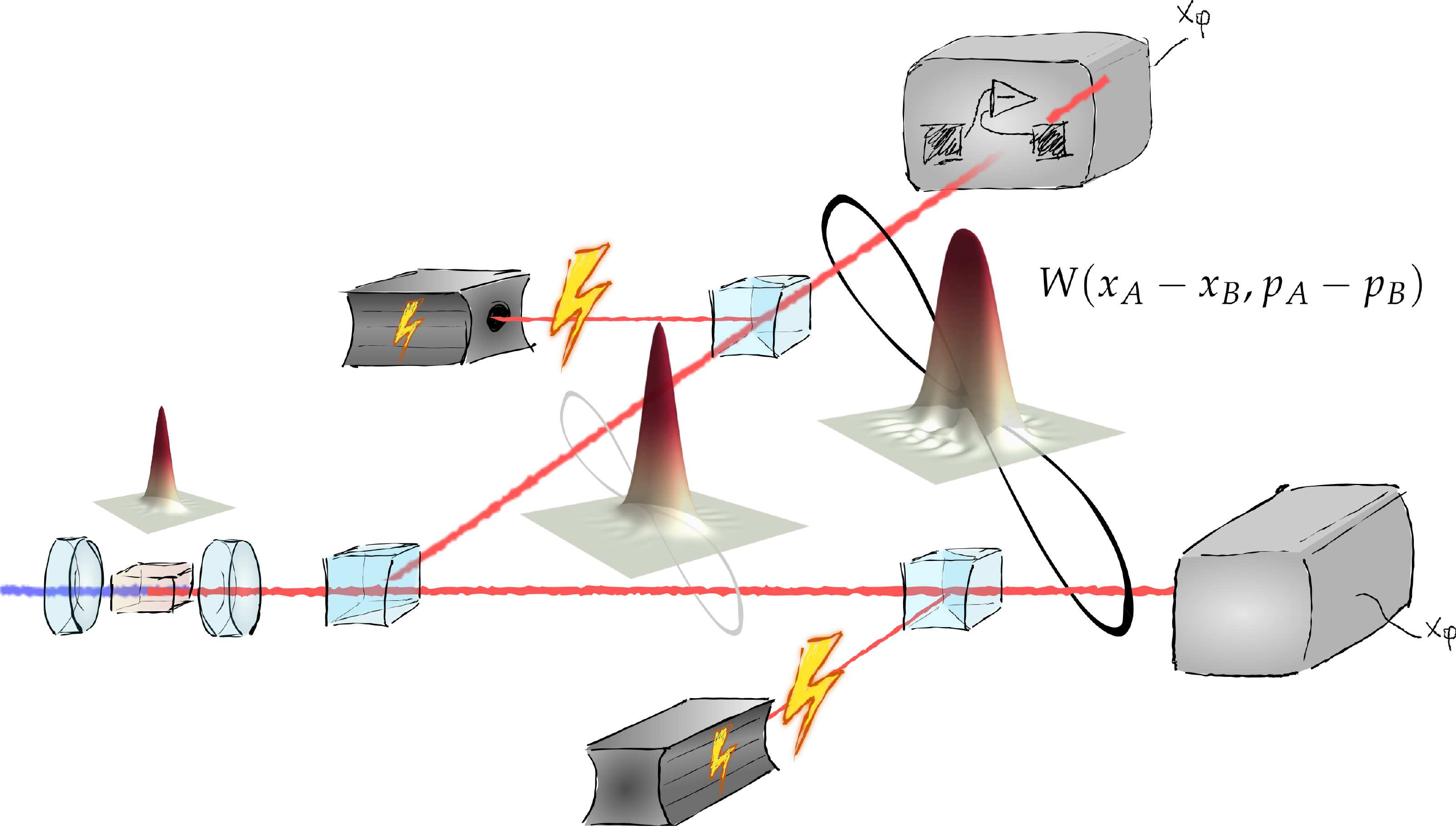}\\
  \caption{Entanglement distillation experimental setup, in this case
  for simultaneous photon subtractions at both Alice and Bob. The
  Wigner functions illustrate the measured and reconstructed states of
  the A-B mode. The A+B mode is in all cases vacuum.
  \label{fig:distillation-setup}}
\end{figure}

In our experiment (Figure \ref{fig:distillation-setup}) we generate a
two-mode entangled state simply by splitting a squeezed vacuum of
variable squeezing level on a 50:50 beamsplitter. This creates a state
with rather weak entanglement, equivalent to that of a standard
two-mode squeezed (EPR) state with half as much squeezing. For Alice
and Bob's local operations, we use the photon subtraction technique
described in the previous section, with a separate channel of frequency
filters and APD for both. A click of the APD detectors signals a
successful distillation event. If Alice and Bob were to proceed with an
application of their shared entanglement, they would have to
classically tell each other at what times they got a detector click. We
do not go that far here -- instead we simply characterize the resulting
quantum state. To quantify the entanglement before and after the
distillation, we reconstruct the joint two-mode density matrix by
homodyne detection on both output modes for both the undistilled state
and for the state with a photon subtracted simultaneously from each of
the two modes. It turns out that even subtracting a photon from just
Alice \emph{or} Bob also increases their shared entanglement, so we did
a third series of measurements of this scenario. The photon subtraction
beamsplitter reflectivities were set at 0\% for the initial state
measurement, at 5\% for the single photon detection series and at 10\%
for the two simultaneous photon detection series (as a compromise
between success rate and entanglement gain).

\begin{figure}
  \includegraphics[width=\columnwidth]{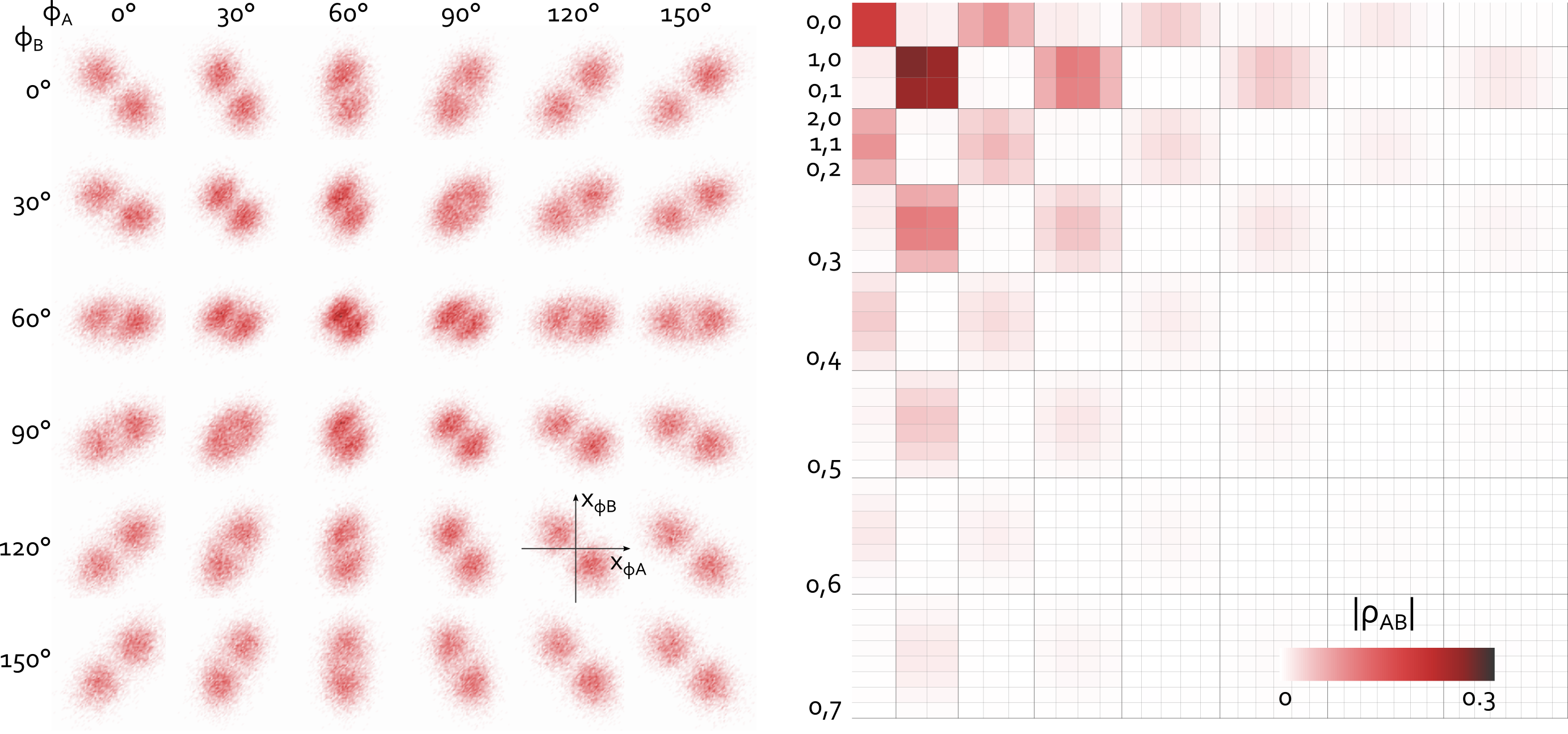}\\
  \caption{Full two-mode homodyne tomography of a single-party
  distilled state. On the left are the joint quadrature distributions
  for all 36 combinations of Alice's and Bob's LO phases. From these
  data the full density matrix (right) was reconstructed. When
  combining the density matrix elements in groups with the same total
  photon number, the overall block structure resembles a single mode
  photon-subtracted squeezed vacuum.
  \label{fig:twomodedm}}
\end{figure}

It is possible to do a complete two-mode tomographic measurement by
independently varying the LO phases of Alice's and Bob's homodyne
detectors. Figure \ref{fig:twomodedm} shows an example of a density
matrix reconstructed from such a measurement, where 36 LO phase
combinations (0-0, 0-$\pi$/6, ..., 0-5$\pi$/6, $\pi$/6-0, ...,
$\pi$/6-5$\pi$/6, ..., 5$\pi$/6-5$\pi$/6) were used. However, this
process is very time-consuming, and due to our particular kind of
initial entangled state, the output state can be separated into a
vacuum state in the $x_A+x_B, p_A+p_B$ variables and a state in the
$x_A-x_B, p_A-p_B$ variables that corresponds to a photon-subtracted
squeezed vacuum. This fact allows us to do the two-mode tomography
using only the 6 phase combinations where the two LO's are in phase, a
strategy that is experimentally justified by confirming that the
virtual `+' mode is indeed in a vacuum state. For details, refer to the
Supplementary Information of \citep{Takahashi2010a}. In broad terms,
the argument relies on the fact that the experimental setup is fully
equivalent to a setup where the photon subtraction (by 1 or 2 APDs) is
done \emph{before} the 50:50 beam splitting. That same fact shows that
for the single-party distillation operation, the joint output state is
independent on whether Alice or Bob detected the photon.

\begin{figure*}
  \includegraphics[width=\textwidth]{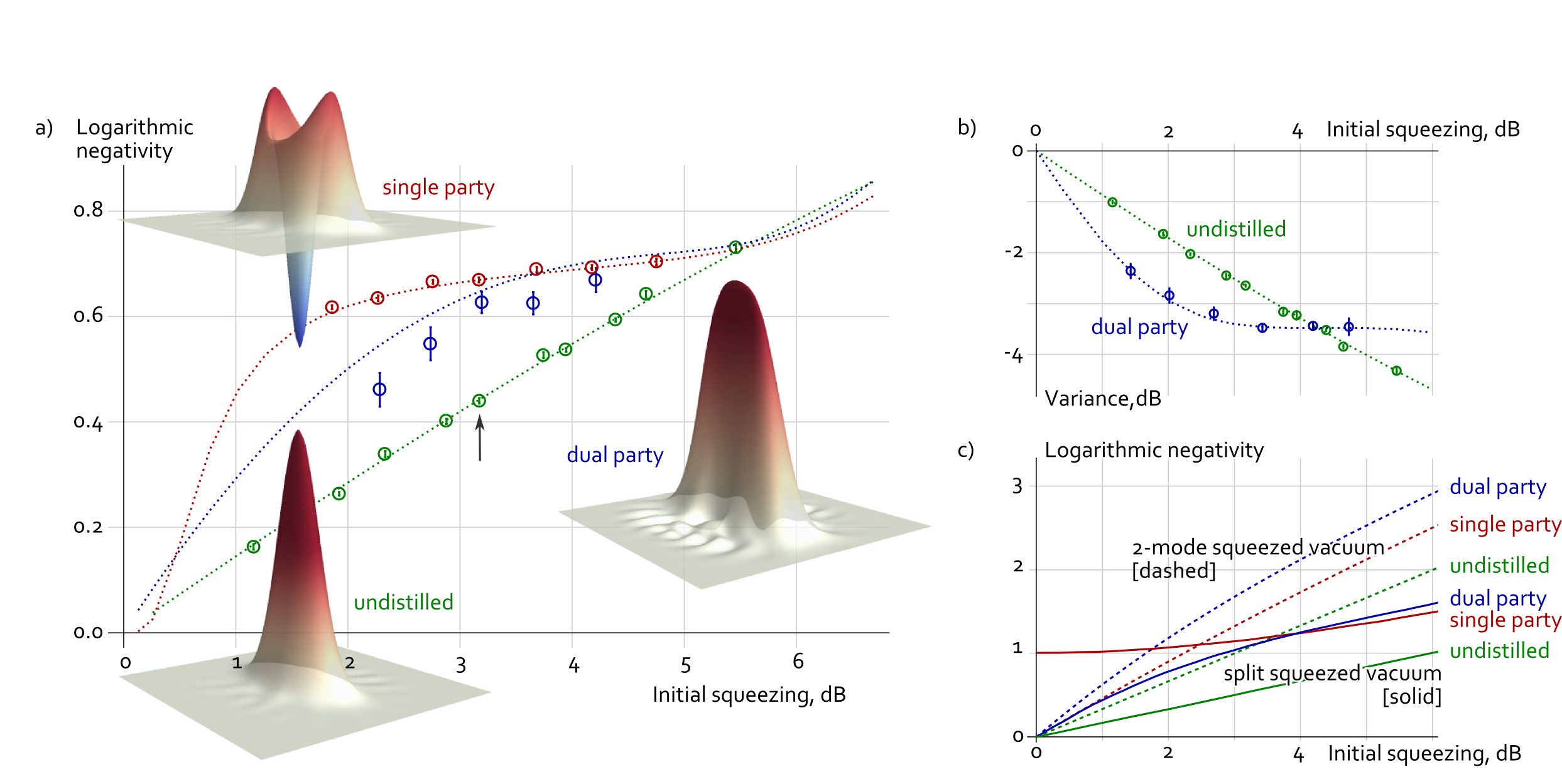}\\
  \caption{Entanglement distillation. \textbf{a)} Logarithmic
  negativity for the undistilled, single-party (1 click) distilled and
  dual-party distilled states for various input squeezing levels.
  Dashed curves are calculated from a model taking all experimental
  parameters into account. The 3D plots are the reconstructed Wigner
  functions $W(x_A-x_B,p_A-p_B)$ for 3.2 dB initial squeezing.
  \textbf{b)} $p_A-p_B$ quadrature variance for the undistilled and the
  dual-party distilled states. Because of losses and other
  imperfections, the undistilled output squeezing is lower than the
  initial level, whereas the distilled state is actually more strongly
  squeezed than the initial state. \textbf{c)} Theoretically calculated
  entanglement negativity for pure states of both 50:50 split squeezed
  vacuum as used in the experiment and of standard two-mode squeezed
  vacuum which initially has twice as much entanglement. Both
  single-party and dual-party photon subtraction are effective for both
  classes of entangled state. \label{fig:distillation-result}}
\end{figure*}

\subsection{Results}

The result of the distillation is shown in Figure
\ref{fig:distillation-result}a. The entanglement measure we use is the
logarithmic negativity \citep{Vidal2002} which is easily computable
from the density matrix. The graph shows that for a wide range of
initial squeezing levels, the entanglement gain from the distillation
protocol is significant for both single-party and dual-party
operations. Perhaps surprisingly, the single-party distillation mostly
works better than the dual-party. This is a result of the particular
kind of entangled initial state we use. It can be understood by
realizing that for weak squeezing levels, the shared state after a
single photon subtraction is essentially a split single photon which
has a full e-bit of entanglement, whereas for two photon subtraction
(or zero), the shared state is mostly vacuum with a small contribution
of photon pairs and therefore with entanglement close to zero. However,
the effect of the single-party distillation operation is not limited to
this particular scenario -- the pure-state calculations in Figure
\ref{fig:distillation-result}c show that also for the more general
two-mode squeezed vacuum state there is a marked increase in
entanglement for both single-party and dual-party photon subtractions.

For a complete Gaussian input to Gaussian output distillation protocol,
the procedure demonstrated here should be expanded to include several
similarly prepared states and then be combined with an iterative
Gaussification procedure \citep{Browne2003}. However, even after just
this single step the resulting distilled state is immediately useful
for e.g. improvement of teleportation fidelity \citep{Opatrny2000,
Cochrane2002, Olivares2003}, at least for the dual-party photon
subtraction. This can be seen from Figure
\ref{fig:distillation-result}b, showing that the two-click distilled
state not only has higher entanglement, but also a higher degree of
squeezing than the input.

%Finally, as shown in \ref{fig:distillation-result}b, in our scenario
%the dual-party operation has the extra advantage that the distilled
%state not only has higher entanglement but also a higher degree of
%squeezing than the input -- something potentially useful for further
%applications such as as a teleportation resource.

\section{Continuous variable qubit states}

The discovery by \citet{Knill2001a} that quantum computation is
possible using only linear optics in combination with single photon
sources and photon detectors gave birth to a wealth of research into
linear optics quantum computation (LOQC). Specifically, a number of
proposals suggested to use coherent states as the basis for
circuit-style LOQC \citep{Cochrane1999a, Jeong2002a, Ralph2003b}; that
is, to encode logical qubits into the physically higher-dimensional
coherent states $|\alpha\rangle$ and $\left|-\alpha\right\rangle$. It
was shown that such schemes could allow efficient computation with only
simple linear operations in-line, with the hard operations put off-line
to the preparation of the resource states. For large amplitudes
$\alpha$ the two states are basically orthogonal, but for small
$\alpha$ they have a finite overlap -- nonetheless, this scheme is
capable of fault-tolerant quantum computing with certain advantages (as
well as disadvantages) over traditional single photon-encoded schemes
\citep{Lund2008a}. One particular advantage is -- in contrast to the
usual two qubit case -- that all four Bell states can be unambiguously
discriminated, leading to higher efficiency for e.g. quantum
teleportation \citep{Jeong2008a}. It is also noteworthy that for
long-distance optical communication, it has been shown that simple
coherent states attain the channel capacity, but the receiver would
have to employ collective decoding with a quantum computer for coherent
state signals to extract the maximum information
\citep{Giovannetti2004}.

%
%Inspired by a number of proposals for coherent state-based quantum
%computing \citep{Cochrane1999a, Jeong2002a, Ralph2003b, Lund2008a}, we
%developed \citep{Takeoka2007} and experimentally demonstrated
%\citep{Neergaard-Nielsen2010a} a method for preparation of complex
%superposition states. The idea of those proposals is to encode logical
%qubits into the physically higher-dimensional coherent states
%$|\alpha\rangle$ and $\left|-\alpha\right\rangle$. For large amplitudes
%$\alpha$ the two states are basically orthogonal, but for small
%$\alpha$ they have a finite overlap -- nonetheless, this scheme is
%capable of fault-tolerant quantum computing with certain advantages (as
%well as disadvantages) over traditional single photon-encoded schemes
%\citep{Lund2008a}.

To realize a coherent state-based computing scheme as described here,
we need access to resources of arbitrary qubits $a|\alpha\rangle +
b\left|-\alpha\right\rangle$. The diagonal states with $a=1, b=\pm1$
(unnormalized) were already approximately realized in the various
kitten state experiments mentioned in the introduction, and of course
the two basis states are trivial, but all other qubit states still
remain to be demonstrated with high fidelities. We suggested a way to
accomplish the generation of these complex coherent state
superpositions using two-photon subtraction of squeezed vacuum with an
added displacement \citep{Takeoka2007}. The displacement before the
second photon detector can be adjusted in phase and amplitude,
according to which the conditional output state is prepared in any
arbitrary superposition of one-photon and two-photon subtracted states
(basically equivalent to a rotation of the coherent state qubit space)
as long as $\alpha$ is small.

\begin{figure}
  \includegraphics[width=\columnwidth]{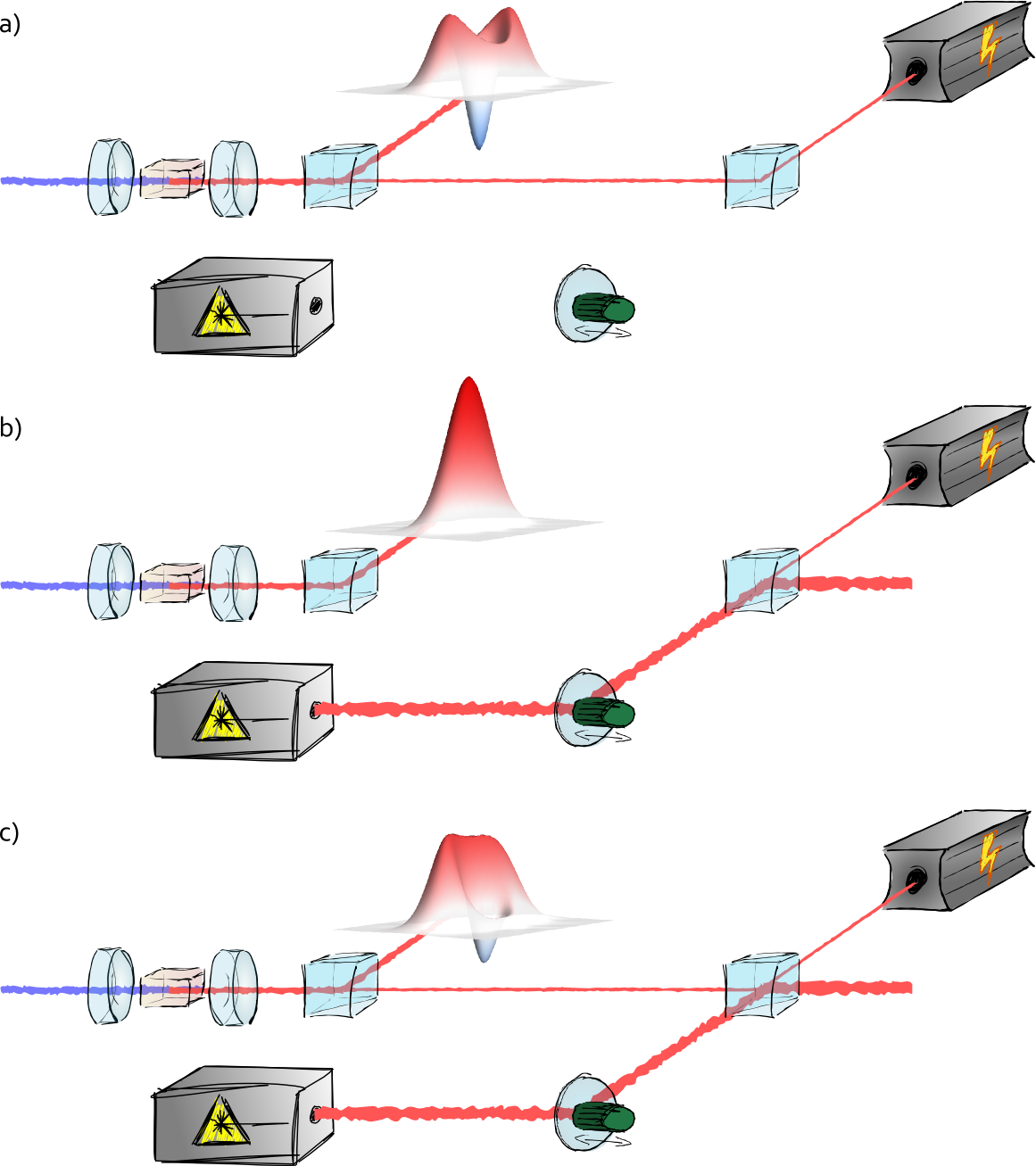}\\
  \caption{Superposition state generation by displaced photon
  subtraction. \textbf{a)} Standard odd kitten state factory.
  \textbf{b)} With the trigger beam from the squeezing blocked, all
  photon clicks are from the uncorrelated auxiliary beam. The output
  state is then just squeezed vacuum. \textbf{c)} With trigger beam and
  auxiliary displacement beam overlapped on a highly reflective
  beamsplitter, the two possible sources of the photon detection are
  indistinguishable and the output is projected to a superposition
  state determined by the relative strength and phase of the two beams.
  \label{fig:qubit-concept}}
\end{figure}

In our experimental study \citep{Neergaard-Nielsen2010a}, we
demonstrated this method of displaced photon subtraction, although we
used only a single photon detector. Our generated states are therefore
superpositions of squeezed vacuum and single-photon subtracted squeezed
vacuum (equivalent to a squeezed photon). The concept is illustrated in
Figure \ref{fig:qubit-concept}. In a standard kitten state generation
setup we insert a highly reflective beamsplitter in the trigger beam
just before the APD. On that beamsplitter we inject a strong coherent
beam with variable phase and amplitude which acts to displace the state
of the trigger beam in phase space. Now the photon detection on the APD
becomes ambiguous; there are two possible, indistinguishable sources of
the detected photon \footnote{The bandwidths of the squeezed vacuum and
the coherent displacement beam are quite different, but the very wide
integrated bandwidth of the APD erases this source of
distinguishability.}, so the total effect of a click on the trigger
beam is a projection onto $\beta|0\rangle+|1\rangle$ and therefore the
conditional output state ends up in a superposition of single-photon
subtracted squeezed vacuum and normal squeezed vacuum. The exact
parameters of the superposition depend on the phase and amplitude of
the displacement beam. If the displacement beam is strong (weak), the
output will be close to a squeezed vacuum (photon).

\begin{figure}
  \includegraphics[width=\columnwidth]{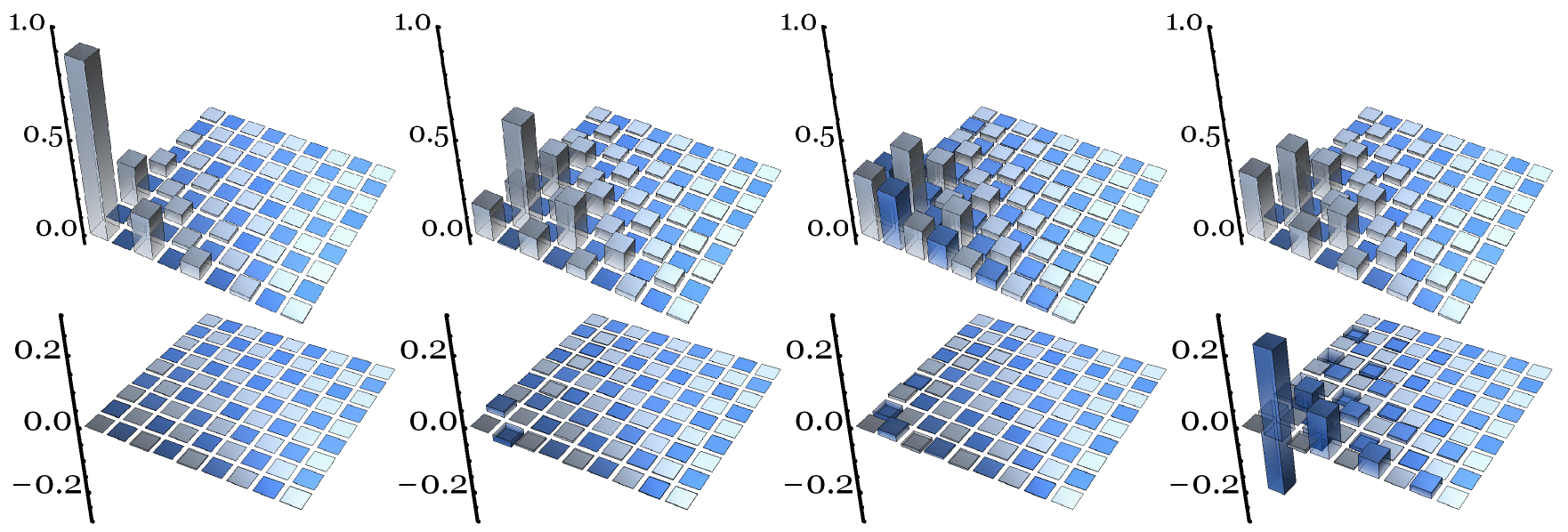}\\
  \caption{Fock state basis density matrices of representative
  reconstructed output states. From left to right: Squeezed vacuum,
  squeezed photon, a state generated with a displacement beam of 1/3
  the intensity (at the APD) of the squeezed trigger beam and with
  180$^{\circ}$ relative phase, and finally a similar state but with
  90$^{\circ}$ relative phase. The upper matrices are real parts, lower
  matrices are imaginary. \label{fig:qubit-dm-normal}}
\end{figure}

To illustrate the potential of this procedure, we carried out a large
number of state preparation and characterizations for different
displacement parameters. Because we restricted ourselves to
single-photon subtraction, we could perform detailed homodyne
tomography (360,000 samples at 12 fixed phases) for all these states
within a reasonable time span. Figure \ref{fig:qubit-dm-normal} shows
the density matrices of just a few of these states: squeezed vacuum,
squeezed photon, and two states roughly halfway in-between where the
trigger and displacement beams were at 180$^{\circ}$ and 90$^{\circ}$
relative phases. No correction for experimental inefficiencies were
performed (and the same goes for the previous distillation results). We
have established a detailed model of the full experiment that fits the
measured outcomes very well without any free parameters, showing that
we have a very good understanding of the individual components of the
setup and providing guides to how to improve the results
\citep{Takeoka2010}.

\begin{figure}
  \includegraphics[width=\columnwidth]{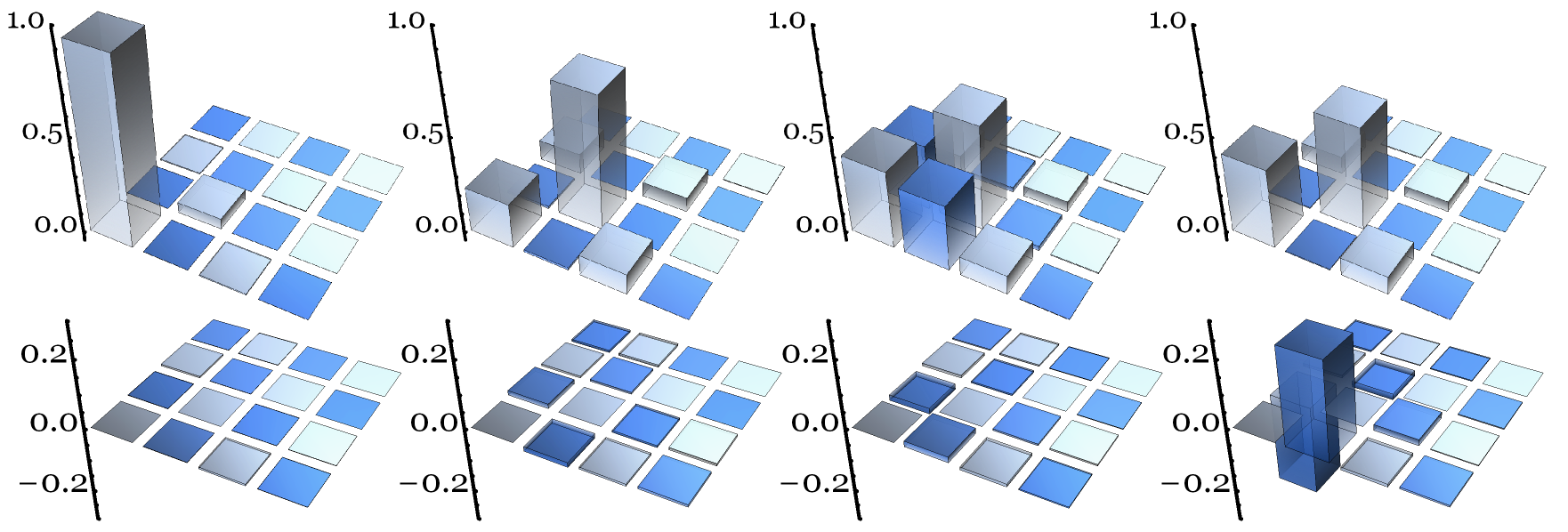}\\
  \caption{Squeezed Fock state basis density matrices of the same
  states as in Figure \ref{fig:qubit-dm-normal}.
  \label{fig:qubit-dm-squeezed}}
\end{figure}

This experiment truly bridges the dv and cv worlds. Not only do we use
experimental methods from both, but the output states seen as a whole
can also be considered as qubits of squeezed states, being
superpositions of the two orthogonal states $\hat{S}(r)|0\rangle$ and
$\hat{S}(r)|1\rangle$. To clearly see this, we have transformed the
density matrices to a different basis in Figure
\ref{fig:qubit-dm-squeezed}, namely the squeezed Fock state basis
$|\phi_k\rangle = \hat{S}(r)|k\rangle$ -- since the squeezing operator
is unitary, this is a proper transformation where the new basis is
orthonormal and complete.
%To clearly see this, we have transformed the
%density matrices to a different basis in Figure
%\ref{fig:qubit-dm-squeezed}, namely the squeezed Fock state basis
%$|\phi_k\rangle = \hat{S}(r)|k\rangle = \sum_n c^k_n(r)|n\rangle$,
%where $c^k_n(r)$ are the Fock state coefficients of the squeezed
%$k$-photon Fock state.
%%\footnote{Since the squeezing operator is
%%unitary, it is easy to see that the squeezed Fock state basis is
%%orthonormal and complete.}.
%%$\langle m|\hat{S}^{\dagger}\hat{S}|n\rangle = \delta_{mn}$, $\sum_n
%%\hat{S}|n\rangle\langle n|\hat{S}^{\dagger} = \mathbb{I}$.).
In this squeezed basis, the states are essentially confined to the
lowest 2-dimensional subspace $\{|\phi_0\rangle,|\phi_1\rangle\}$, with
a small contribution from higher-order states mostly due to
experimental inefficiencies. From these density matrices, we can
calculate the parameters for each state when interpreted as a squeezed
qubit
\begin{equation}
\left|\psi\right\rangle = \cos\frac{\theta}{2}\hat{S}(r)|0\rangle +
e^{i\phi}\sin\frac{\theta}{2}\hat{S}(r)|1\rangle .
\end{equation}
For the two non-trivial states of Figures \ref{fig:qubit-dm-normal},
\ref{fig:qubit-dm-squeezed} the parameters are $\theta = 100^{\circ}$
and $\phi = 0^{\circ}, 90^{\circ}$, respectively. The states are not
pure -- they range between 0.56 to 0.88 in purity -- and therefore the
squeezed vacuum and squeezed photon are also not perfectly orthogonal,
with an overlap of 0.26. The generated qubit states are therefore not
directly usable for quantum information purposes as such, but they
still serve as a clear demonstration of the potential of the displaced
photon subtraction method. In fact, it is also applicable to genuine
cat states, since the photon subtraction operation $\hat{a}$ turns the
even cat state $|\alpha\rangle + \left|-\alpha\right\rangle$ into the
odd $|\alpha\rangle - \left|-\alpha\right\rangle$ and vice versa, and
the displaced photon subtraction can therefore generate an arbitrary
superposition of these two states. It was also pointed out by
\citet{Marek2010a} that the operation corresponds to a single-mode
phase gate in the coherent state basis, and that in more complex
settings, it can be used to implement the Hadamard gate and a two-mode
phase gate, which just serves to emphasize its generally useful
quality.

\begin{figure*}
  \includegraphics[width=\textwidth]{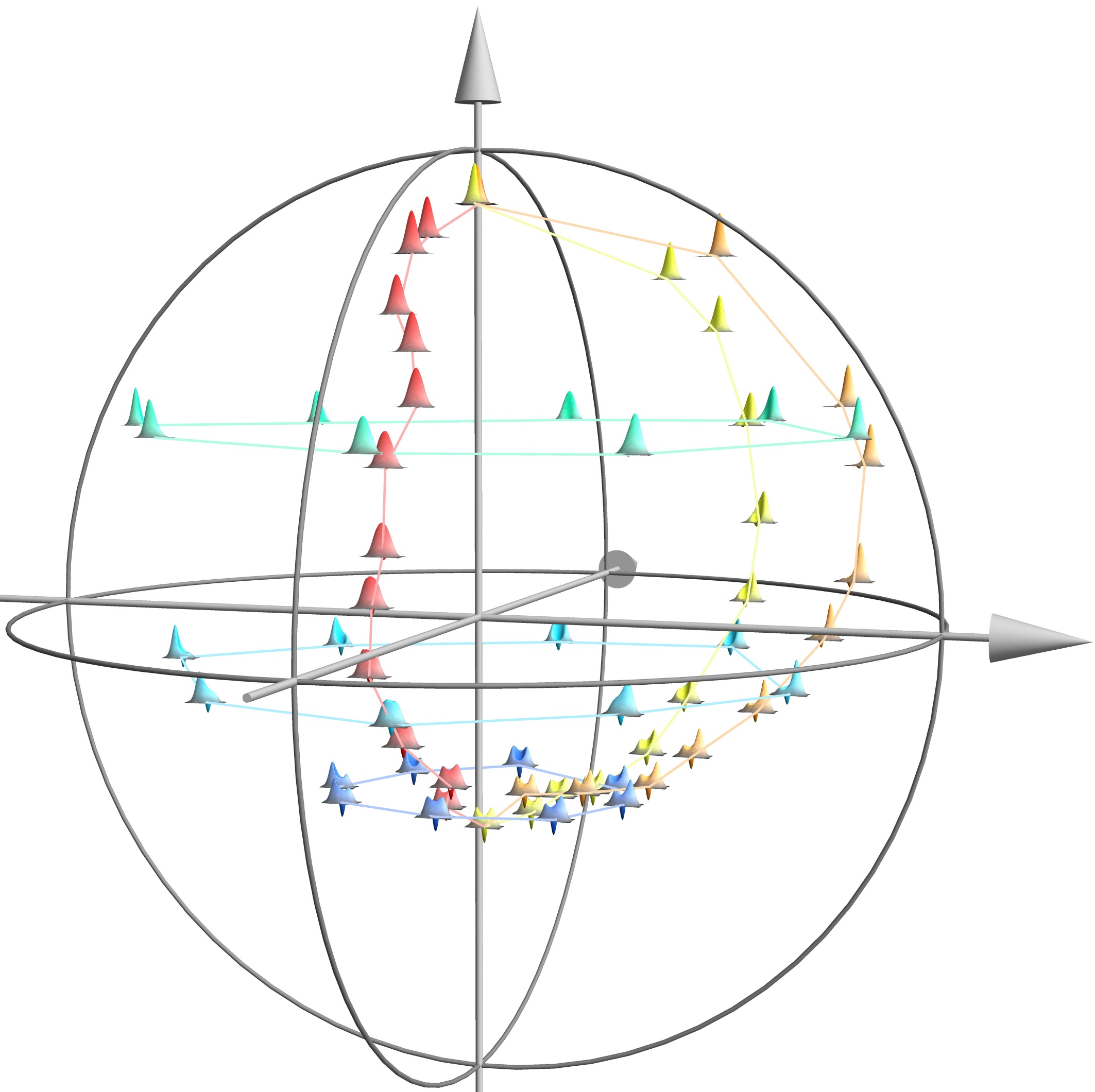}\\
  \caption{Bloch sphere with the experimentally generated and
  reconstructed qubit states plotted by their Wigner functions. The
  north pole basis state is $\hat{S}(r)|0\rangle$ and south is
  $\hat{S}(r)|1\rangle$, both with $r=0.38$. \label{fig:bloch}}
\end{figure*}

To finish this section and to emphasize the dv--cv link, we present in
Figure \ref{fig:bloch} the full set of generated qubit states,
represented by their reconstructed Wigner functions (cv) and inserted
at their appropriate positions in a Bloch/Poincar{\'e} sphere (dv)
according to the qubit parameters extracted from their squeezed basis
density matrices. To see the detailed parameters and close-up Wigner
functions of each state, please go to our group's website where an
interactive Java applet is available \citep{QubitApplet}.

\section{Conclusion}

We have presented two different examples of what can be accomplished
with the photon subtraction operation on non-classical states of light.
From a practical applications point of view, it enables a number of
different quantum information protocols in continuous variable or
hybrid regimes. But it also highlights that the division between the
discrete variable and continuous variable worlds is not as large as it
was previously made out to be, perhaps to the extent that it hardly
makes sense to distinguish them. Judged from the amount of activity on
these kinds of hybrid processes and schemes, both on the theoretical
and experimental side, we expect to see many more breakthroughs in the
near future, ultimately leading to practically applicable quantum
information technologies.

\bibliography{icqit}

\end{document}